\documentclass[twocolumn,prb,showpacs,10pt]{revtex4-1}

\usepackage{graphicx}
\usepackage{epstopdf}

\usepackage{color}

\begin{document}

\title{Field-induced Conductance Switching by Charge-state Alternation
in Organometallic Single-Molecule Junctions}

\author{Florian Schwarz}
\affiliation{IBM Research - Zurich,
S\"{a}umerstrasse 4, 8803 R\"{u}schlikon, Switzerland}
\author{Georg Kastlunger}
\affiliation{University of Vienna,
Department of Physical Chemistry, Sensengasse 8/7, 1090 Vienna,
Austria} \affiliation{TU Wien - Vienna University of Technology, Institute
for Theoretical Physics, Wiedner Hauptstrasse 8-10, 1040 Vienna,
Austria}
\author{Franziska Lissel} \affiliation{Department of Chemistry,
University of Z\"{u}rich, Winterthurerstrasse 190, 8057
Z\"{u}rich, Switzerland}
\author{Carolina Egler-Lucas} \affiliation{Department of Chemistry,
University of Z\"{u}rich, Winterthurerstrasse 190, 8057
Z\"{u}rich, Switzerland}
\author{Sergey N.\ Semenov} \affiliation{Department of Chemistry, University of Z\"{u}rich,
Winterthurerstrasse 190, 8057 Z\"{u}rich, Switzerland}
\author{Koushik Venkatesan}
\affiliation{Department of Chemistry, University of Z\"{u}rich,
Winterthurerstrasse 190, 8057 Z\"{u}rich, Switzerland}
\author{Heinz Berke}
\affiliation{Department of Chemistry, University of Z\"{u}rich,
Winterthurerstrasse 190, 8057 Z\"{u}rich, Switzerland}
\author{Robert Stadler}
\affiliation{University of Vienna,
Department of Physical Chemistry, Sensengasse 8/7, 1090 Vienna,
Austria} \affiliation{TU Wien - Vienna University of Technology, Institute
for Theoretical Physics, Wiedner Hauptstrasse 8-10, 1040 Vienna,
Austria}
\author{Emanuel\ L\"{o}rtscher}
\affiliation{IBM Research - Zurich,
S\"{a}umerstrasse 4, 8803 R\"{u}schlikon, Switzerland}
\email{eml@zurich.ibm.com,robert.stadler@tuwien.ac.at,hberke@chem.uzh.ch}

\begin{abstract}

Charge transport through single molecules can be
influenced by the charge and spin states of redox-active metal
centres placed in the transport pathway.\ These molecular
intrinsic properties are usually addressed by varying the
molecule's electrochemical and magnetic environment, a procedure
that requires complex
 setups with multiple terminals.\ Here we show that oxidation and reduction of
organometallic compounds containing either Fe, Ru or Mo centres
can solely be triggered by the electric field applied to a
two-terminal molecular junction.\ Whereas all compounds exhibit
bias-dependent hysteresis, the Mo-containing compound additionally
shows an abrupt voltage-induced conductance switching, yielding
high-to-low current ratios exceeding 1000 at voltage stimuli of
less than 1.0 V.\ DFT calculations identify a localized,
redox-active molecular orbital that is weakly coupled to the
electrodes and closely aligned with the Fermi energy of the leads
because of the spin-polarised ground state unique to the Mo
centre.\ This situation opens an additional slow and incoherent
hopping channel for transport, triggering a transient charging
effect of the entire molecule and a strong hysteresis with
unprecedented high low-to-high current ratios.\\
\end{abstract}

\maketitle

Switching an electric signal from a low- to a high-current state
is one of the key elements in an electric circuit with
applications in signal processing, logic data manipulation or
storage.\ In current Si-based technology with device dimensions
approaching the sub-5nm range, it becomes increasingly difficult
to maintain large high-to-low ratios mainly because of leakage
currents.\ Therefore alternative switching mechanisms are needed.\
In single-molecule electronics, a variety of intrinsic
conductance-switching mechanisms\cite{Molen2010} exist: Gating of
the molecular orbitals (MOs) by electrostatic\cite{Song2009} or
electrochemical means\cite{Liao2010}, which requires a third
electrode, or modifying specific photoactive molecular structures,
e.g.\ by optically irradiating the molecule to form or break
bonds\cite{Irie2000,Dulic2003,Kronemeijer2008,vanderMolen2009}.\
Also mechanically-induced changes in the molecule--metal
coupling\cite{Quek2009} can lead to conductance alternations.\
Another trigger is the electric field inherently present in a
molecular transport junction: Conformational changes due to
interactions between the electric field and molecular dipoles were
demonstrated to alternate the conductance of single-molecule
junctions\cite{Blum2005,Loertscher2006a,Meded2009} by up to a
factor
of 70.\\

Potentially more powerful mechanisms exploit intrinsic molecular
quantum phenomena related to spin and charge states.\ An early
example of addressing the spin state of a single
molecule\cite{Kahn1998,Sato2007,Baadji2009} revealed Kondo
resonances using cobalt (Co) metal centres.\cite{Park2002} More
recently, a spin cross-over was induced by an electric field in
iron (Fe)-based molecular nanoclusters\cite{Prins2011a} (with
high-to-low ratios of $\sim$ 2), and in a coupled spin pair of two
Co atoms\cite{Wagner2013} (with high-to-low ratios of 2 - 3).\
Regarding intrinsic charge states, Coulomb blockade peaks were
reported in ruthenium (Ru)-containing wires\cite{Kim2007}, but not
confirmed in self-assembled monolayers.\cite{Luo2011} On the
single-molecule level, Ru-based molecules showed
conformation-induced changes in the conductance\cite{Ruben2008}
rather than changes due to intrinsic redox mechanisms.\ Two Ru
metal centres in a photochromatic compound demonstrated reversible
light-induced conductance switching in ensemble
junctions.\cite{Meng2012,Meng2014} In another study, the
importance of the copper (Cu) coordination on the conductance was
demonstrated\cite{Ponce2014}.\\

\noindent \textbf{Placing Individual Metal Centres in the Transport Pathway}\\

Earlier we studied dinuclear organometallic Fe compounds with
various anchoring schemes\cite{Schwarz2014a,Lissel2014} and
discovered indications of field-induced conductance switching in
the case of weak molecule--metal coupling.\ Motivated by these
findings, we have developed mononuclear organometallic compounds
of the type (MeCOSC$_6$H$_4$-C$\equiv$C-)M(P$\cap$P)$_2$ (M = Fe;
P$\cap$P = 1,2 - bis(diethylphosphino)ethane:\ (\textbf{1}); M =
Ru, Mo (Molybdenum); P$\cap$P = 1,2
-bis(diphenylphosphino)ethane:\ (\textbf{2}), (\textbf{3})) using
weak thiol coupling\cite{Reed1997,Zotti2010} to preserve
molecule-internal spin and charge degrees of freedom for the
solid-state molecular junctions.\ Fe, Ru and Mo were chosen as
metal centres.\ The synthetic strategy aims at placing the metal
centres right in the transport pathway (Fig.\ 1a,b) to achieve an
optimal influence on transport and maximum interaction with the
electric field.\ We used identical acetylenic backbones to
constrain the variable parameters to the metal centres and their
ligand fields.\ To prevent dimerization, the sulphur end groups
were acetyl-protected, with the protection groups being hydrolyzed
\emph{in situ}, forming the metal--molecule--metal junctions
Au--\textbf{1$'$}--Au, Au--\textbf{2$'$}--Au, and
Au--\textbf{3$'$}--Au (Fig.\ 1c).\ For the Fe metal centre, we
used the bidentate phosphine ligands depe (depe =
1,2-bis(diethylphosphino)ethane) and for the Ru\cite{Wuttke2014}
and Mo centres, dppe (dppe = 1,2-bis(diphenylphosphino)ethane)
chelate ligands.\ To extend the molecular length to 2.5 nm (S--S
distance), we chose phenylene spacers as conducting backbones.\
The synthesis of \textbf{2} was reproduced using a previously
reported procedure\cite{Touchard1997,Wuttke2014}, whereas new
synthetic protocols were established for \textbf{1} and \textbf{3}
(see SI for details).

\noindent \textbf{Conductance Switching in Single-Molecule Junctions}\\

First, we perform current--voltage $I$--$V$ data acquisition by
repeatedly forming and breaking the junction\cite{Loertscher2007}
(see SI).\ In the entire data sets, we find a substantial number
of curves ($\sim$ 90 $\%$) that all exhibit distinct features
differing from conventional non-linear molecular transport,
namely, curves with hysteretic behaviour.\ Here, the curves
acquired for sweeps from negative to positive bias are separated
in a given voltage range from those acquired in the opposite
direction.\ For the Au--\textbf{1$'$}--Au junction, around 85 $\%$
of the curves show hysteresis, for Au--\textbf{2$'$}--Au 80 $\%$,
and for Au--\textbf{3$'$}--Au 95 $\%$.\ Fig.\ 1d shows 50
representative $I$--$V$ curves taken at 50 K (see SI sections 10,
16-18 for statistics, sampling rate and temperature dependence).\
The hysteretic behaviour of the three compounds differs in the
voltage range and the transition between the two envelopes.\
Accordingly, we categorised the $I$--$V$s into two types: Type
\emph{I} curves are found for all compounds and are characterised
by a small hysteresis that affects only a particular section of
the voltage (blue backgrounds in Fig.\ 1d), whereas the $I$ --$V$s
for the low- and high-bias regimes are nominally identical.\ The
conductance gap (as defined by the onset in transport) is not
altered, and the transition between the curves is continuous.\
Type \emph{II} curves are only found for the Mo compound and
differ from type \emph{I} curves by an approximately 100x lower
current and an abrupt switching between two distinct curves,
accompanied by a hysteresis.\ Here, the conductance gaps change
substantially (e.g.\ from 0.15 V to 0.85 V).\ Fig.\ 1e summarises
the experiments schematically by providing also the sweep
directions.\ When analysing the occurrences of type \emph{I} and
type \emph{II} curves, we find that they depend on the junction
configuration: Type \emph{II} curves are found just before
breaking the molecular junction and show a switching between two
distinct states (Fig.\ 2a).\ When extracting the maximum
high-to-low ratio in the hysteresis region and plotting it versus
the corresponding voltage (Fig.\ 2b), we find that type \emph{I}
curves display a narrow energy distribution, whereas type
\emph{II} curves seem to depend non-linearly on energy, with an
increasing ratio for increasing bias.\ The high-to-low current
ratios are 1.5 to 20 for type \emph{I} switching consistently for
all compounds, and exceed 1000 for type \emph{II} switching.\\

\noindent \textbf{Coherent Tunneling and Decoherent Hopping Transport}\\

In principle, several possible explanations exist for the
hysteretic curves at smaller junction distances (type \emph{I})
and the abrupt switching at larger junction distances (type
\emph{II}):\ As the ground state of the Mo compound \textbf{3} and
some of the excited states of the Fe compound \textbf{1} and the
Ru compound \textbf{2} are magnetic, a high-spin/low-spin (HS/LS)
crossover comes to mind, given the observed switching in similar
systems.\cite{Prins2011a,Wagner2013} There, the HS/LS crossover
caused a drastic change in the electronic structure across the
entire electron wavelength spectrum that accompanied the switching
between two distinct electronic states.\ In our experimental data,
however, we do not find such drastic changes, e.g.\  no change in
the conductance gap for type \emph{I} curves and almost identical
curves for certain voltage regimes for both types.\ Moreover, the
nuclei of the molecules would adapt to that and thereby preclude
hysteresis simply because there is nothing that could cause a time
delay.\ In contrast, an oxidation or reduction of the
transition-metal compounds
 would enable a potential observation of hysteresis, as proposed
 theoretically.\cite{Galperin2005,Kuznetsov2007,Migliore2013}
 Those authors argued that if the
charging rate is similar to the bias sweeping rate, a time delay
required for a memory effect could occur, making the charging
visible in the $I$--$V$ curves.\ Here, the conductance is governed
by a coherent tunnelling channel mediated by a delocalized
molecular orbital (MO) (``fast channel"), whereas hysteresis is
related to charging of a localized MO in an electron-hopping
channel (``slow channel") (Fig.\ 3a).\ The probability of the
localized MO to be occupied determines the respective conductance
contributions from the two charging states at every bias
increase.\

To simulate transport through single molecules, these charging
probabilities were implemented into a stochastic approach.\ For
calculating $I$--$V$ curves, we combine proposed algorithms
\cite{Migliore2013} with data from density functional theory (DFT)
calculations for the transmission (defining coherent tunnelling)
and the transfer integral, reorganization energy and driving force
(describing electron hopping~\cite{Kastlunger2013, Kastlunger2014,
Kastlunger2015}).\ First, we demonstrate that by varying the ratio
between the bias sweeping rate and the charging/hopping rate, the
abrupt switching shown in Fig.\ 1c can be qualitatively reproduced
for a simple two-MO tight-binding model.\ In Fig.\ 3b, we show
$I$--$V$ curves simulated for forward and reverse bias sweeps for
various coupling strengths, $\gamma$, and fixed voltage sweeping
rates, $u$. For the strongest coupling, $\gamma/u$ = 10$^{12}$
V$^{-1}$, a statistical average of multiple switching events is
observed, and as a consequence, the forward and backward sweeps
fall together with the average of the two limiting $I$--$V$ curves
obtained from the integration of the transmission functions of the
reduced and the oxidized systems (shown as orange and gray dotted
lines in Fig.\ 3b, respectively).\ When lowering the coupling to
$\gamma/u$ = 10$^4$ V$^{-1}$, averaging covers a smaller number of
redox processes per integration time, $\Delta t$, resulting in
fringes.\ For $\gamma$/$u$ = 100 V$^{-1}$, there is roughly one
jump in each integration interval, and for the weakest coupling,
$\gamma$/$u$ = 0.5 V$^{-1}$, only a single jump happens during a
full sweep.\ Going from the strongest to the weakest coupling step
by step (Fig. 3b), the voltage range where both sweeps follow the
lower curve for low bias and the upper one for high bias becomes
larger because ever higher voltages are needed to increase the
likelihood of jumps.\ In the wide range of couplings from
$\gamma$/$u$ = 10$^{12}$ to 100 V$^{-1}$, however, the forward and
backward $I$--$V$ curves still follow the same path, although our
stochastic approach creates uncertainties or line-thickening for
ratios of $\gamma$/$u$ smaller than 10$^8$ V$^{-1}$.\ Only with
$\gamma$/$u$ as low as 1 V$^{-1}$ an irreversible switching or a
``lock-in" process does take place, where the forward sweep
follows the lower limiting curve for the reduced state and the
backward sweep the upper one for the oxidized state; a scenario
that qualitatively explains the type \emph{II} curves for the Mo
junction.\ Fig.\ 3c displays the hopping rates for oxidation and
reduction, showing that the lower the ratio, the later $R_{ox}$
and $R_{red}$ cross the horizontal line defining the sweeping
rate.\ $R_{ox}$ determines the probability of charging the
compounds, whereas the stability of this oxidized state depends
inversely on $R_{red}$.\ Therefore, as can be seen from the
left-hand-side panel, the bias necessary for reaching the charged
state is inversely proportional to the coupling strength.\
$R_{red}$, in contrast, decreases with the coupling strength,
which makes a reduction even at lower biases less likely, finally
enabling the occurrence of a ``lock-in" process during bias
sweeps.\

Let us now look at the electronic structures of the transport
junctions as obtained from DFT calculations (Fig.\ 4).\ Energetic
positions of the MO's and their spatial distributions as well as
the corresponding transmission functions are computed by a
Nonequilibrium Green's Function (NEGF) DFT
formalism\cite{Brandbyge2002,Xue2002,Rocha2005} with the GPAW
code\cite{Mortensen2005,Enkovaara2010}.\ Because the Mo compound
\textbf{3} is the only compound among the three with a
spin-polarised ground state, we show its MO eigenenergies and
transmission functions for spin-up and spin-down separately (green
curves in Fig.\ 4a).\ The magnetic property of the Mo system is
the reason why a very localised MO with $d_{xy}$ symmetry on the
metal atoms (where $z$ is the transport direction) moves close to
the Fermi level for one spin orientation (violet dots).\ In
contrast, this MO lies far outside of any reasonable bias window
for the Fe compound \textbf{1} and the Ru compound \textbf{2}.\ As
a high degree of localization of a MO results in a very weak
coupling to the electrodes, this MO can be considered the ``slow
channel" for the Mo compound, while for the Fe and Ru compounds
the HOMO-1 ($d_{yz}$) plays this role.\ For all three systems, the
``fast channel" is provided by the delocalized HOMO $d_{xz}$
(Fig.\ 4b).\

To calculate the hopping rates for the oxidation/reduction that
govern the switching between neutral and charged compounds, we
follow an approach developed earlier.\cite{Kastlunger2015} Table 1
lists all relevant parameters for both charging states for all
three systems at the equilibrium distance.\ For the Mo compound,
it also gives those values at an elongation of the bonding
distance between the anchor group and the electrodes by 0.5 \AA{}
on both sides according to the experimental findings that type
\emph{II} curves are found for elongated junctions just before
rupture.\ The driving force $\Delta G_{0}$, which is defined by
the energy difference of the ``slow-channel" MO and $E_F$, is
lower for Mo than for Fe and Ru by a factor of 2 - 3.\ Its
transfer integral is two orders of magnitudes smaller and even
three orders of magnitude smaller at the elongated distance.\
Because of the self-interaction problem of DFT, which becomes more
severe for localized states, the calculations overestimate the
spatial extension of the respective orbital and thereby also the
transfer integral.\ Additionally, we have to account for the fact
that the binding of the molecule to the metal surfaces is
idealised in our DFT calculations, where we use perfectly planar
Au(111) surfaces and symmetric bonding of the compounds at
equilibrium distances.\ To account for these aspects, all
calculated transfer integrals are consistently scaled down by a
factor of 100 for the calculated $I$--$V$ curves in the lower
panels of Fig.\ 4c.\ In all panels, different sweeping rates are
used for the forward and backward sweep, in agreement with the
experimental situation (see SI).\ Whereas for the
Au--\textbf{1$'$}--Au and the Au--\textbf{2$'$}--Au junctions, an
elongated configuration reveals only a minor influence on the
hysteresis and the functional behaviour, the Au--\textbf{3$'$}--Au
junction shows an abrupt transition at the weaker coupling
conditions induced by elongation.\ This situation perfectly
reproduces the experimental findings in terms of switching energy,
relative current levels, type of hysteresis and drastic change in
the conductance gap.\ Furthermore, DFT calculations can also
reproduce the high-to-low current ratios, which are around 1.5 -
4.5 for type \emph{I} hysteresis and around 200 for type \emph{II}
hysteresis with abrupt switching (Fig.\ 4c).\

\begin{table}
\footnotesize
 \begin{tabular}{l|c|c|c|c|c}
  & $\Delta G_0$    & $\lambda_{in}$  & $\lambda_{img}$ & $\lambda_{tot}$ & V     \\
  \hline
  Fe (Au--\textbf{1$'$}--Au) & 0.408  &  0.079  &  -0.021   &  0.057  &  $2.6 \times 10^{-3}$ \\
  Ru (Au--\textbf{2$'$}--Au) & 0.499  &  0.094  &  -0.022   &  0.072  &  $2.3 \times 10^{-3}$ \\
  Mo (Au--\textbf{3$'$}--Au) & 0.203  &  0.075   &  -0.017  &  0.058  &  $1.2 \times 10^{-5}$ \\
  Mo (+0.5 \AA{}) & 0.269  &  0.075   &  -0.013  &  0.062  &  $1.3 \times 10^{-6}$ \\
 \end{tabular}
 \caption{Parameters for electron hopping calculated from DFT for the three compounds
 at equilibrium geometry and for the Mo compound \textbf{3} also for a junction with the S-Au bond
 elongated by 0.5 \AA{} on each side. All values are given in eV.}
 \label{hop_table}
\end{table}

\noindent \textbf{Conclusion and outlook}\\

In summary, we have experimentally and theoretically investigated
the transport properties of organometallic molecules containing
Fe, Ru and Mo metal centres in their transport pathway.\ We find
hysteretic transport properties with continuous transitions for
all three transport junctions, and additionally an abrupt
switching for the Mo compound.\ Comprehensive DFT modelling,
taking into account bias-driven charging, indicates an
oxidation/reduction mechanism mediated by a weakly coupled,
localized MO that is unique to the Mo compound because of its
spin-polarized ground state.\ This MO gives rise to abrupt
switching with high-to-low current ratios of more than 1000,
outperforming all previously explored molecular-intrinsic
conductance-switching mechanisms, such as
magnetoresistance.\cite{Schmaus2011} DFT combined with a
two-channel transport model qualitatively agrees with experiments
regarding the functional behaviour of the hysteresis.\ We
therefore conclude that intrinsic redox functionality is
maintained in weakly-coupled solid-state organometallic junctions,
remains accessible at feasible electric fields in a two-terminal
geometry, and can be controlled by tuning the voltage sweeping
rate in respect to the intrinsic oxidation and reduction rates.\
Moreover, by bias-induced charge-state alternations, a conductance
switching with technologically relevant high-to-low current ratios
exceeding 1000 at voltages of 1.0 V could be achieved in a
single-molecule building block.\ Even though technological
parameters, such as fatigue, switching speed, non-volatility etc.,
remain to be determined in real device geometries, such ultimately
scaled building blocks fulfill in principle the requirements for
future memory in terms of reasonably low
operational fields, speed, and large high-to-low current ratios.\\

\providecommand{\latin}[1]{#1}
\providecommand*\mcitethebibliography{\thebibliography} \csname
@ifundefined\endcsname{endmcitethebibliography}
  {\let\endmcitethebibliography\endthebibliography}{}

\noindent \textbf{Acknowledgements}\\

\noindent \footnotesize{We are grateful to M.\ Koch for support
with the synthesis of the end groups and to O.\ Blacque for
single-crystal X-ray diffraction.\ We also acknowledge G.\
Puebla-Hellmann, V.\ Schmidt, and F.\ Evers for scientific
discussions, and M.\ Tschudy, U.\ Drechsler and Ch.\ Rettner for
technical assistance.\ We thank W.\ Riess and B.\ Michel for
continuous support.\ Funding from the National Research Program
``Smart Materials" (NRP 62, grant 406240-126142) of the Swiss
National Science Foundation (SNSF) and the University of
Z\"{u}rich is gratefully acknowledged.\ G.K.\ and R.S.\ are
currently supported by the Austrian Science Fund FWF, project
Nos.\ P22548 and P27272, and are deeply indebted to the Vienna
Scientific Cluster VSC, on whose computing facilities all DFT
calculations were performed (project No.\ 70174). In addition,
G.K.\ receives a grant co-sponsored by the Austrian Academy of
Science \"{O}AW, the Springer Verlag
and the Austrian Chemical Society G\"{O}CH.}\\

\noindent \textbf{Author contribution}\\

\noindent \footnotesize{F.\ L. and G.\ K. made equal contributions
to this work and should therefore be considered joint first authors.
F.\ L., C.\ E., S.\ N.\ S., K.\ V., and
H.\ B.\ designed and synthesized the compounds.\ F.\ S., and E.\
L. set up and performed the experiments and the data analysis.\
G.\ K. and R.\ S.\ carried out the calculations.\ F.\ S., G.\ K.,
K.\ V., H.\ B., R.\ S.\ and E.\ L.\ wrote the paper.\ All authors
discussed the
results and commented on the manuscript.} \\

\noindent \textbf{Additional information}\\

\noindent \footnotesize{Supplementary information accompanies this
paper at www.nature.com/ naturenanotechnology.\ Reprints and
permission information is available online at
http://www.nature.com/reprints/.\ Correspondence and requests for
materials should be addressed to:  hberke@chem.uzh.ch (H.B.),
venkatesan.koushik@chem.uzh.ch (K.V.) for chemistry,
robert.stadler@tuwien.ac.at (R.S.) for DFT calculations, and
eml@zurich.ibm.com (E.L.) for experimental work.\ CCDC-1040144
(for \textbf{1}), CCDC-1040145 (for \textbf{2}) and CCDC-1040146
(for \textbf{3}) contain the supplementary crystallographic data
(excluding structure factors) for this paper.\ These data can be
obtained free of charge from The Cambridge Crystallographic Data
Centre via \emph{https://summary.ccdc.cam.ac.uk/structure-summary-form}}.\\

\noindent \textbf{Competing financial interests}\\

\noindent \footnotesize{The authors declare no competing financial
interests.}\

\noindent \textbf{Chemical Synthesis.} The synthetic steps and
full characterisation of all compounds can be found in the
supporting information.\\

\noindent \textbf{Transport measurements.}
Electron-beam-structured break junctions are mechanically actuated
in a three-point bending mechanism operated under
ultra-high-vacuum conditions (UHV; pressure $p<$ 2 $\times$
10$^{-9}$ mbar) at 50 K.$^{30}$ Molecules are deposited from a
highly diluted solution in dry tetrahydrofuran (THF; 4 $\times$
10$^{-5}$ m/L).\ Electrical characterization is carried out with a
Hewlett-Packard Semiconductor Parameter Analyzer HP4156B upon
repeated opening and closing of the molecular junction (more
details can be found in the supporting information).\\

\noindent \textbf{Computational details.} All calculations of
transmission probabilities $T(E)$ and $I$--$V$ curves were
performed within a NEGF-DFT framework$^{37-39}$ with the GPAW
code.$^{40,41}$ We chose a linear combination of atomic orbitals
(LCAO) on a Double Zeta level with polarisation functions (DZP)
for the basis set and a Perdew-Burke-Ernzerhof (PBE)
parametrisation for the exchange-correlation (XC) functional.\ The
MO eigenenergies were calculated by decoupling the basis functions
localised on the molecule from those of the surface states via a
subdiagonalisation of the transport Hamiltonian.$^{34}$

For the redox process, we combine a recent formalism$^{33}$ with a
coherent tunnelling description based on NEGF-DFT for the
calculation of the $I$--$V$ characteristics of the reduced and
oxidized states and a hopping description of the redox reaction
based on Marcus theory$^{36}$.\ By calculating the bias-dependent
reaction rates of oxidation and reduction, a probability $P$ can
be determined that describes the system's probability to be in one
of the respective charge states after a given integration time
$\Delta t$.\ To simulate single $I$--$V$ sweeps, we apply a
stochastic approach, in which we trap the system into one distinct
charge state in every step.\ By calculating the change of
probability $dP$, defined by either $dP=R_{ox}dt$ or
$dP=R_{red}dt$, between two time steps $t$ and $t+dt$, where
$dt\ll\Delta t$, and comparing $dP$ defined in this way with a
random number between 0 and 1, we create a criterion for the
switching between the two states.\ The overall current is then
calculated from a mean value $I(V)=\frac{1}{n}\sum_{i=1}^nI(V,i)$,
averaging over all $n$ current values, with $\Delta t=n*dt$.\ More
details can be found in the supplementary information.

\begin{figure*}[htbp]
    \begin{center}
    \resizebox{2.0\columnwidth}{!}{\includegraphics{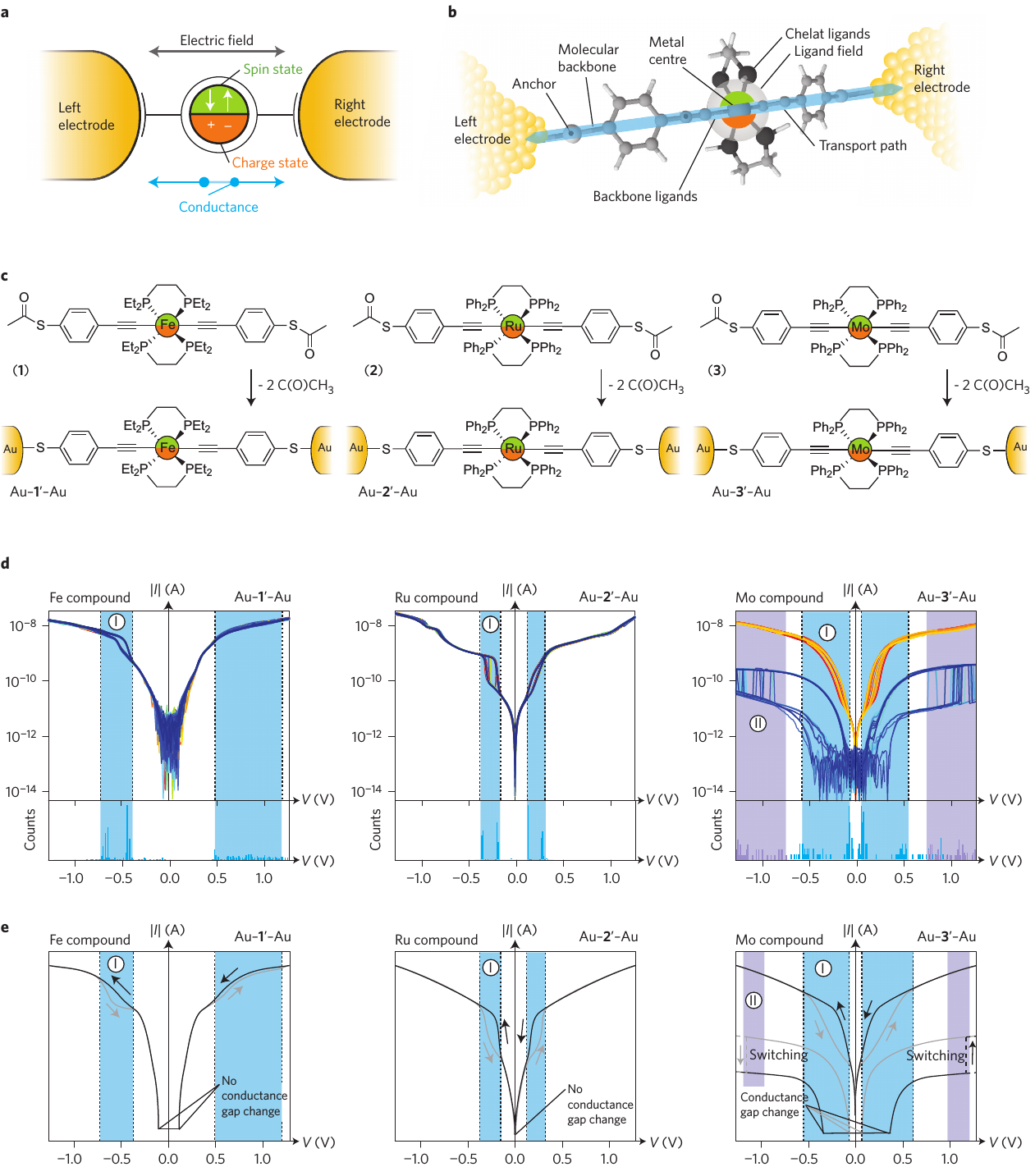}}
    \caption{\label{fig:1} \textbf{Organometallic single-molecule junctions bearing Ru, Fe and Mo metal centres to
    provide charge and spin degrees of freedom.} \textbf{a}, Addressing spin (green) and charge
    (orange) states of a
    molecular junction by the electric field present by means of the bias
    applied in a two-terminal
    geometry.\ \textbf{b}, Anatomy of the organometallic molecular junction
    with one metal centre
    placed directly in the charge-transport pathway.\
    \textbf{c}, Mononuclear compounds of type \emph{trans}-(MeCOSC$_6$H$_4$-C$\equiv$C-)M(P$\cap$P)$_2$
    (M = Fe; P$\cap$P = 1,2
- bis(diethylphosphino)ethane: (\textbf{1}); M = Ru, Mo; P$\cap$P
= 1,2 -bis(diphenylphosphino)ethane: (\textbf{2}), (\textbf{3})).\
Shown are also the representative transport junctions
Au--\textbf{1$'$}--Au, Au--\textbf{2$'$}--Au, and
Au--\textbf{3$'$}--Au, respectively.\
    \textbf{d}, 50 representative $I$--$V$ characteristics taken at 50 K:\
    The Fe \textbf{1} and Ru compounds \textbf{2} both show a continuous splitting of
    the $I$--$V$s, providing a hysteresis region
    (blue background).\ Here, no change in the conductance gap or in
    the high-bias conductance is found.\ The same type of curves (labelled \emph{I})
    is found also for the Mo compound \textbf{3} with a slightly larger hysteresis regime.\
    In addition,
    there is a second type of curves (labelled \emph{II}) that reveals an
    abrupt switching accompanied by a
    large change in the
    conductance gap (violet background).\ \textbf{e}, Schematic representation of
    the two types of hysteresis with continuous \emph{I} (Fe, Ru and Mo compounds
    \textbf{1}, \textbf{2}, and \textbf{3})
     and abrupt switching \emph{II} (Mo compound \textbf{3} only).
    }
\end{center}
\end{figure*}
\newpage

\begin{figure*}[htbp]
    \begin{center}
    \resizebox{1.5\columnwidth}{!}{\includegraphics{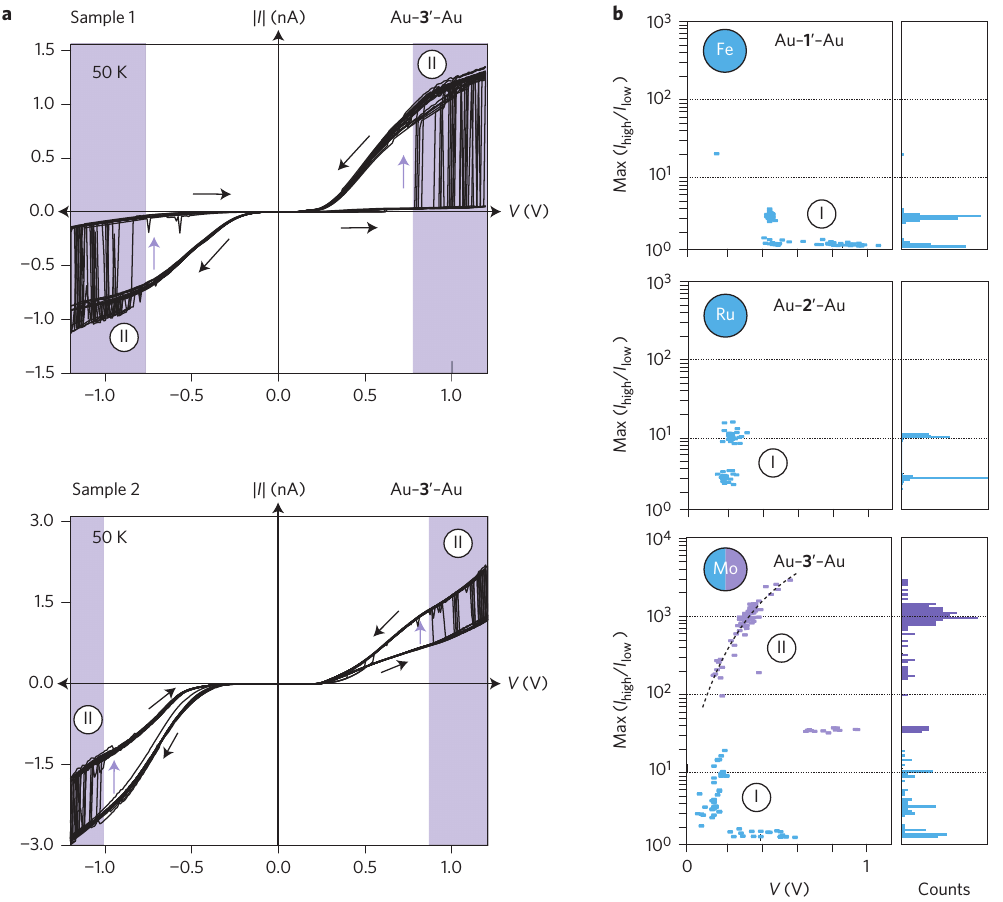}}
    \caption{\label{fig:2} \textbf{High-to-low current ratios for type \emph{I} and type \emph{II} curves}. \textbf{a},
     Type \emph{II} hysteretic curves on
    linear current scale
    found for the Mo junction Au--\textbf{3$'$}--Au, revealing an abrupt switching from a low- to
    a high-conductance regime (50 K).\ Plotted are two data sets
    measured on fully independent samples with typical differences
    for single-molecule experiments regarding the current amplitude and switching voltages.\
    \textbf{b}, Statistics on maximum high-to-low current ratios extracted in the hysteresis regime for type \emph{I} and
    type \emph{II} curves for all compounds.\ Whereas type
    \emph{I} ratios vary from 1.5 (Fe compound \textbf{1}) to 20
    (Mo compound \textbf{3}), the type \emph{II}
    switching found only for the Mo compound \textbf{3} attains 2500 at a bias of less
    than 0.6 V.\ The ratio seems to follow a voltage dependence as indicated by the black dotted line.}
\end{center}
\end{figure*}
\newpage

\begin{figure*}[htbp]
    \begin{center}
    \resizebox{2.0\columnwidth}{!}{\includegraphics{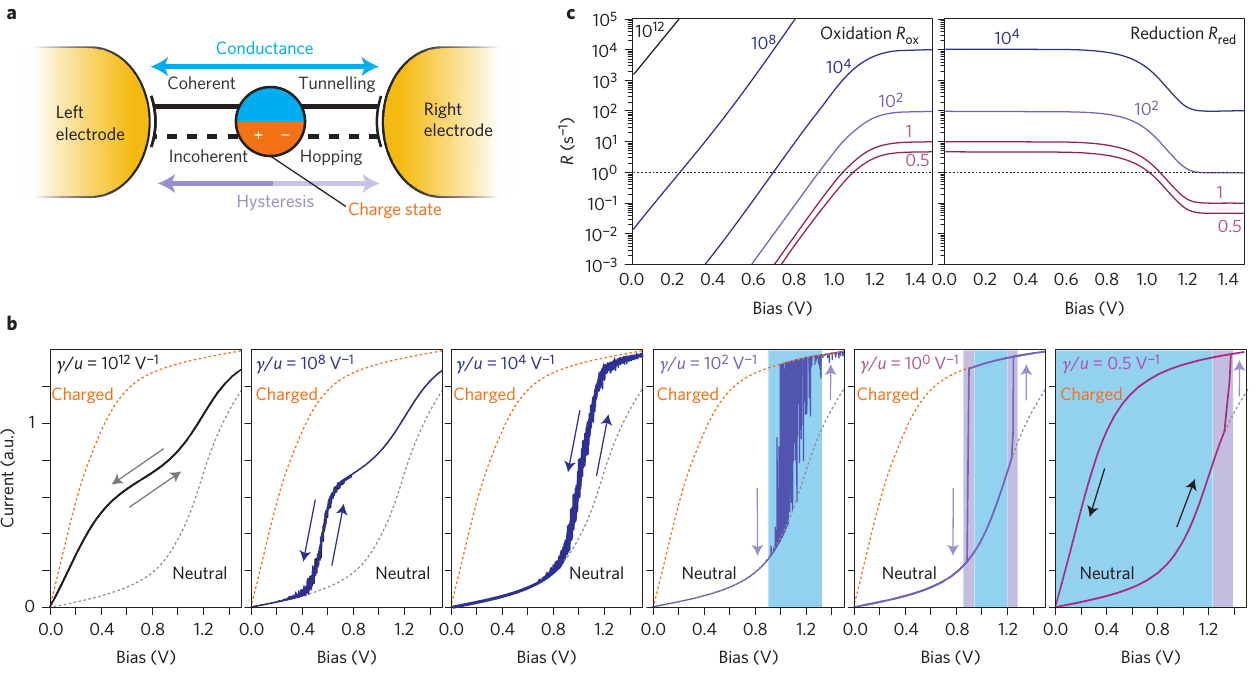}}
    \caption{\label{fig:3} \textbf{Two-channel transport mechanism in single-molecule junctions and
    its influence on current switching and hysteresis}. \textbf{a}, Model for a two-channel transport mechanism through
    a molecular junction with
    a fast coherent tunnelling path (cyan) responsible for the conductance, and
    a slow, incoherent hopping path (violet) causing hysteresis, e.g., by charging.\
    \textbf{b}, Calculated $I$--$V$ curves from a 2-state tight-binding model, with one
    strongly and one weakly coupled
    molecular orbital (MO).\
    The parameters used in this model are $\Delta
    E_{ox/red}$ = 0.1 eV for the localized MO, a reorganization energy
    of 0.1 eV and a bias sweep rate of 1 V/s; the onsite
    energies for the delocalized MO in the neutral and the charged system
    are set to -0.6 eV and -0.1 eV, respectively.\ The driving force
    $\Delta G_{0}$ of the reaction depends on both $\Delta E_{ox/red}$
    and the average bridge population in the fast channel.\cite{Migliore2013}
    In our calculations, the bias sweeping rate determines the integration
    time, $\Delta$$t$, between two voltages, whereas the hopping rate for
    both oxidation and reduction is defined by the square of the same
    MO-metal-coupling $\Gamma$, or the super exchange
    rates $\gamma=\frac{2\pi}{\hbar}\Gamma^2$, which thereby determines the
    number of hopping processes from the reduced to the oxidized state and
    vice versa that occur within $\Delta$t.\
    The current follows
    a monotonous trace between neutral (dotted line) and charged
    (orange line) state for $\gamma/u$ $\geq$ $10^{4}$ V$^{-1}$, but the $I$--$V$ characteristics reveals
    a bistable range for decreasing rates that finally leads to hysteresis (blue background) and an abrupt switching
    for rates around 1 (violet background indicates abrupt switching regimes).\
    \textbf{c}, Corresponding reaction rates for oxidation, $R_{ox}$, and reduction, $R_{red}$, as a function of bias and different
    ratios of the coupling strength, $\gamma$, and the bias sweeping rate, $u$.}
\end{center}
\end{figure*}
\newpage

\begin{figure*}[htbp]
    \begin{center}
    \resizebox{2.0\columnwidth}{!}{\includegraphics{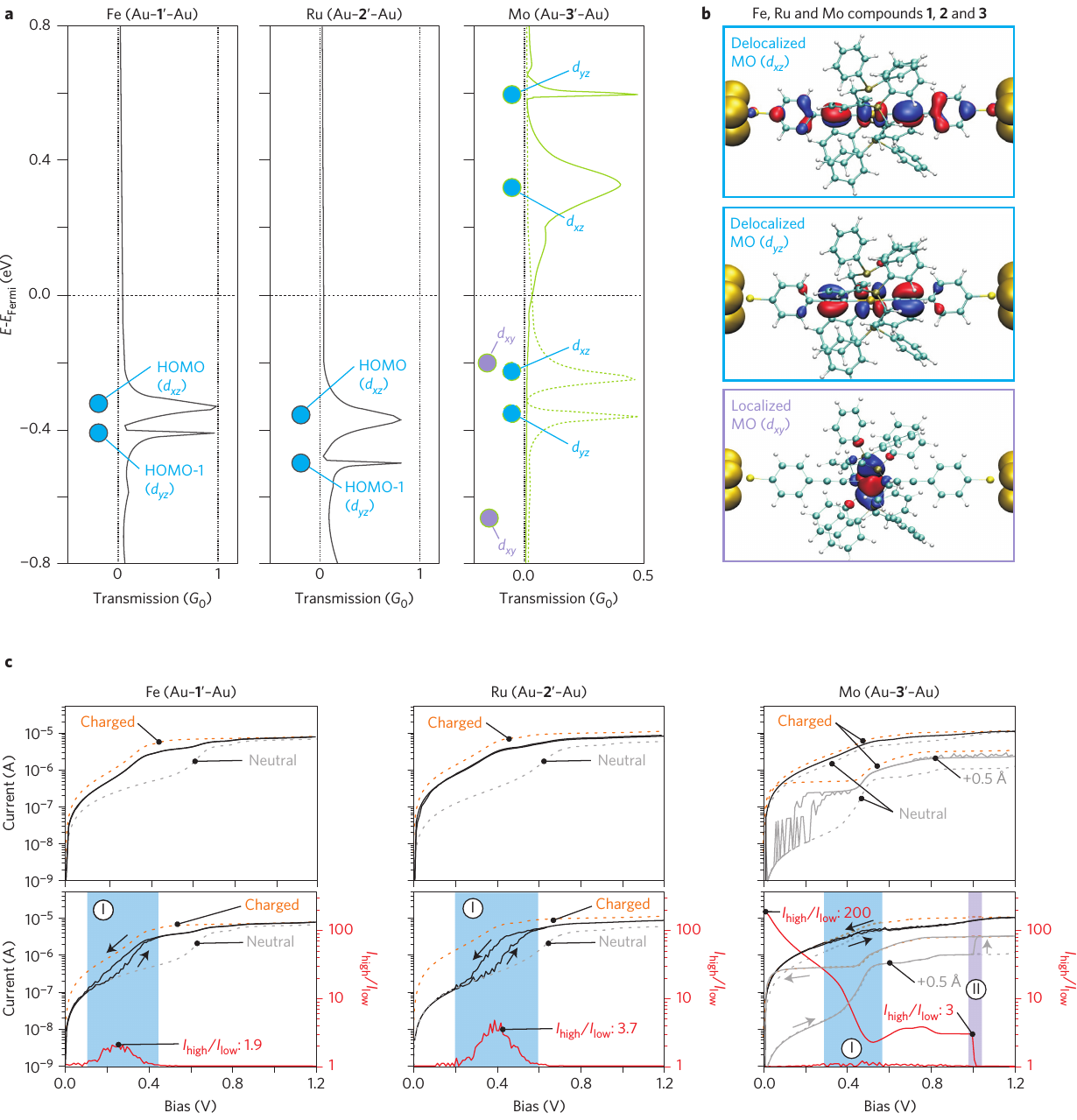}}
    \caption{\label{fig:4} \textbf{DFT-derived transmission and molecular orbitals as well as
    transport characteristics calculated under finite bias}. \textbf{a},
    Total transmission functions (grey and green curves) and selected MO eigenenergies
    (dots) with
    respect to $E_{F,Au}$
    obtained from a sub-diagonalization
    of the transport Hamiltonian, and
    the respective transmission functions calculated with
    NEGF-DFT; the two are plotted separately for spin-up and
    spin-down electrons for the Mo compound \textbf{3}
   because of the latter's spin-polarized ground state (green lines).\
    \textbf{b}, Spatial distributions of the
    MOs with the delocalized HOMO and HOMO-1 and a localized MO (with
   $d_{xy}$
    symmetry) found only for the Mo compound to lie in the relevant energy
    window.\
    \textbf{c}, $I$--$V$ curves calculated by a NEGF-DFT approach
 combined with a hopping description for
 the charging of the slow channel\cite{Kastlunger2015} by using the
 hysteresis formalism\cite{Migliore2013}.\
The upper
 three panels show uncorrected $I$--$V$ curves using the parameters
 given in Table 1, whereas for the calculations in the lower panels,
the transfer integral is scaled down by a factor of 100 for reasons given in
 the main text.\ Only for the Mo compound \textbf{3} a weaker coupling results in the
 abrupt switching curves.\ The red curves provide the ratio between high and
 low currents for the two types of hysteresis.\ All $I$--$V$ curves were simulated
 from 100 bias steps in
 each direction with integration
  times of 0.01 s for the forward and 0.1 s for the backward sweeps.}
\end{center}
\end{figure*}

\end{document}